# Autonomy with regard to an Attribute

Eric Sanchis
*Laboratoire de Gestion et Cognition*
*Université Paul Sabatier*
*115 route de Narbonne*
*31077 Toulouse, France*
sanchis@iut-rodez.fr

**Abstract**

*This paper presents a model of autonomy called autonomy with regard to an attribute applicable to cognitive and not cognitive artificial agents. Three criteria (global / partial, social / nonsocial, absolute / relative) are defined and used to describe the main characteristics of this type of autonomy. A software agent autonomous with regard to the mobility illustrates a possible implementation of this model.*

## 1. Introduction

Since the second half of the Nineties, *autonomous agents* are used more and more in the design and in the implementation of complex computer systems. The association of the concepts of agent and autonomy was explored and commented on in many theoretical works, primarily in the MultiAgent Systems (MAS) area. A first point of view which was developed by researchers such as Jennings and Wooldridge, or Franklin and Graesser was to consider autonomy as a global property, i.e. a property which applies to the agent in its totality.

For Jennings and Wooldridge [7], "*an agent is a computer system situated in some environment, and that is capable of autonomous action in this environment in order to meet its design objectives*". The *situated* aspect means that the agent perceives its environment and that it is able to modify it. *Autonomy* means that the agent is able to act without the direct intervention of human beings or other agents, and that it is able to control its own actions and its internal state. For the authors, autonomous agents are not an innovation because systems like software daemons (ex: xbiff) or a simple thermostat can be classified as autonomous agents.

Franklin and Graesser [6] synthesize their point of view on the agents with the following definition: "*an autonomous agent is a system situated within and a part of an environment that senses that environment and acts on it, over time, in pursuit of its own agenda and so as to effect what it senses in the future*". For the authors, the software agents are present in many fields but any software program is not an agent: for example, according to their definition a payroll program cannot be considered as an agent. On the contrary, a thermostat fully satisfies their definition of an agent, i.e. an entity provided with autonomy.

The main purpose of the autonomy model that we propose is to remove a major defect present in the two models outlined previously: the *global* aspect of autonomy. Indeed, a problem of global autonomy is that it is carrying paradox and confusion [8]. Let us illustrate this problem with an example. The definition of an autonomous agent as *an agent able to act without the direct intervention of other agents and having a control on its own actions and its internal state* is not entirely satisfactory. One can notice that even the human agents do not have this control neither on their internal state nor on all the actions they carry out. However, a (partial) autonomy of a human agent cannot be seriously questioned. This apparent paradox seems to result from an insufficient decomposition of the agent.

Moreover, as many researchers had already noticed, autonomy is not an ordinary property but a complex property. More precisely, our work led us to distinguish two categories of properties of deeply different nature: *qualities* and *attributes* [10]. *Qualities* characterize properties with vague and elastic contours, difficult (perhaps impossible) to have an intimate knowledge. They are properties which content changes according to the point of view which one can have. They authorize multiple modelizations and different interpretations, when these interpretations are not contradictory. Some properties which can be identified as qualities are *autonomy* and *intelligence*. Attributes are properties intrinsically less difficult to encircle than qualities. Examples of attributes are *mobility*, *replication* or *perception*. They are generally reduced to a mechanism with one or more well defined procedures. Contrary to qualities, it is always possible to say if an agent has or does not have a specific attribute. Attributes play a significant role in the type of autonomy which we propose.

Lastly, in the MAS field several models of autonomy were designed for goal-directed cognitive agents. For such an agent, a goal corresponds to a cognitive representation which an agent has of the world in which it is immersed. It results that these models of autonomy are not appropriate to non cognitive agents. The model of artificial autonomy described in section 4 is sufficiently general to apply to cognitive and non cognitive agents.

The paper is structured as follows. Section 2 presents the main characteristics of the autonomy models generally associated to artificial agents. In the next section, the significant features of these models are used to define a set of criteria which will be used to compare these models and to introduce the autonomy with regard to an attribute. Lastly, section 4 will describe the theoretical and practical aspects of this model and some possible extensions.

## 2. Artificial Autonomy approaches

Autonomy is a typical example of what we have called a *quality*. Indeed, autonomy characteristics seem difficult to be synthesized with only one theory or model. In fact, reducing autonomy to a single model removes the essence of this property. This is why many models of autonomy were proposed by researchers in the Artificial Intelligence area. Various aspects can be brought out from these models: the *organic autonomy* of biological inspiration (Varela, Bourgine), the *social autonomy* based on the power relationships between agents (Castelfranchi, Scerri) and the *decisional autonomy* founded on the choice of the agent (Barber, Vendryès), only to quote the most representative types of autonomy.

The autonomy models briefly outlined below partially illustrate the significant disparity of the points of view generated by this property.

The works of Luck and D'Inverno [9] on autonomy is integrated into a more general theory on the interactions between agents being based on the concepts of *goal* and *motivation*. The authors distinguish two very different points of view about autonomy: autonomy as the possibility for the agent to generate its own goals and autonomy as relations of dependence of an agent upon others. According to the first point of view, autonomous agents generate their own goals from their own motivations. According to the second point of view, the autonomy of an agent is modulated by the existence of relations of dependence between agents. This distinction separates what we will call thereafter *social autonomy* and *nonsocial autonomy*.

Barber and al. [1], [2], consider autonomy as the possibility for an agent to influence the decision-making process in the resolution of a given problem. Four levels of autonomy are defined. First level: the agent carries out the orders which are provided to it by another agent. Second level: the agent collaborates with other agents in order to achieve a common goal. Third level: the agent plans and carries out its own actions. Lastly, on the fourth level, the agent plans the actions for itself and for other agents. The agent is completely autonomous when it completely manages the decision-making process. It is partially autonomous when it shares this decision-making with other agents. Finally an agent which does not take part in this process is heteronomous.

Castelfranchi and Falcone [3], [4], connect the concept of autonomy to the concept of *dependence*. They explain why on the one hand, autonomy is a relational concept - relation between an active entity and its environment or other active entities (social autonomy) -, and why on the other hand this relation also derives from the properties of the internal architecture of the entity.

Adjustable autonomy describes the property of an autonomous system to change its level of autonomy among several levels while the system continues to function. A level of autonomy indicates a particular distribution between manual operations and automatic operations [5]. The distribution of autonomy dynamically changes in order to optimize the global performances of the system [12].

## 3. Criteria for a partial classification

In order to compare in a uniform manner the autonomy with regard to an attribute and the preceding models, three criteria of analysis were defined:
- Autonomy as a *global* or *partial* property of the agent
- Autonomy as a *social* or *nonsocial* property
- Autonomy as an *absolute* or *relative* property.

**Global or partial autonomy.** Autonomy is said *global* when the definition applies to the entire agent. In the field of MAS, researchers generally use the expression "*autonomous agent*" [6] [7]. Another definition which typically illustrates this global aspect of autonomy is: *autonomy is the condition of a person or a community who determines by itself the law to which it is subjected*. In this definition, this quality also applies to the entity considered (agent, individual and group) in its totality. On the other hand, the nature of the law determined by the entity is not specified. We interpret this general information in the following way: as soon as an agent is able to create a law, it is globally autonomous. By replacing the concept of law by the concept of goal, we can qualify the model presented in [9] as *global autonomy*.

Autonomy is *partial* when the property does not apply any more to the entire agent but to a part of it. The agent is in this case autonomous with regard to something. The model presented in [2] can be described as an example of *partial* autonomy because the autonomy is defined for each agent goal and the agent can be non autonomous (heteronomous) with regard to a specific goal.

**Social or nonsocial autonomy.** Autonomy is described as *social* when its definition or characterization explicitly refers to one (or several) other(s) agent(s). It is *nonsocial* in the contrary case.

The (X, Y, B) triplet defined in [4] perfectly characterizes *social* autonomy: the autonomy of agent *X* from agent *Y* about goal *B*. The presence of agent *Y* is essential to the various variations of this model (collaborative autonomy, non collaborative autonomy, etc). Models which use a decision-making process shared by several agents, as defined in adjustable autonomy [12] or in [2] are also and naturally models of *social* autonomy. The agents considered in the models of *social* autonomy are not of unspecified nature but are *cognitive* agents, i.e. provided with a certain type of intelligence.

The autonomy presented in [9] is *nonsocial* because it is based on the possibility of an agent to generate a goal without implying the presence of another agent. However, it should be noted that a model of autonomy described as *nonsocial* does not mean that this model is inapplicable when several agents are in relation. This qualifier simply means that the definition of the model does not impose the presence of another agent.

**Absolute or relative autonomy.** Autonomy is *absolute* when only one level of autonomy is defined: the agent (or part of the agent) is autonomous or is not autonomous [9].

Autonomy is *relative* when several *levels* of autonomy are defined. The interest is to be able to define a measurement of the autonomy of an agent. In [2], the status of the agent can pass from simple executant to decision maker.

It must be noticed that we distinguish carefully *types of autonomy* and *levels of autonomy*. A *type of autonomy* underlines the nature of this one: autonomy of goals, autonomy of plan, autonomy from the environment, etc. Our work on autonomy concerns a specific type of autonomy: autonomy with regard to an attribute.

**Summary.** The preceding criteria are useful
- To analyze existing models: the autonomy of an agent presented in [2] is *partial* (because defined for each goal), *relative* (several levels of autonomy were identified) and *social* (some relations of dependence exist between the agents). The model developed in [9] is a *global* autonomy model (the agent is able or not able to create a goal, its nature not being specified), *absolute* (only one level is defined) and *nonsocial* (the presence of another agent is not essential), and
- To design new models: we developed a model of autonomy called *autonomy with regard to an attribute*. This autonomy is *partial*, *absolute* and *nonsocial*. It is (1) *partial*, because it applies only to a part of the agent: a specific attribute, (2) *absolute*: the agent is autonomous or is not autonomous with regard to this attribute, (3) *nonsocial*: the model does not require the presence of a second agent (Table 1).

**Table 1. Partial taxonomy of some autonomy models.**

|  | *global* | *partial* | *social* | *nonsocial* | *absolute* | *relative* |
|---|---|---|---|---|---|---|
| Barber |  | X | X |  |  | X |
| Luck | X |  |  | X | X |  |
| Sanchis |  | X |  | X | X |  |

## 4. Autonomy with regard to an Attribute

Autonomy with regard to an attribute is a model of decisional autonomy. It is stated in the following way: an agent is autonomous with regard to an attribute A if it can choose in a nondeterministic way a policy of use p among several and if it can change this one during its execution time.

An *attribute* corresponds to a simple and non ambiguous property of an agent. A *policy of use* means in our context a succession of actions which carries out a particular and well identified functioning mode. Let us illustrate the concepts of *attribute* and *use policies* with two examples: *replication* and *mobility*.

*Replication* is an attribute which makes it possible for an agent to create a clone of itself on the local or a remote host. The diversity of replication policies can (among other possibilities) come from the replication rate (one and only one clone per site, not more than one clone per site, at least one clone per site) and from replication hosts (replication on all the sites accessible to the agent, sites meeting a specific criterion).

Another example of attribute is *mobility*. It is this attribute which will be used to illustrate the various aspects of the model.

### 4.2. Model

Two elements characterize autonomy with regard to an attribute:
- A set of operating modes of this attribute
- A two-component choice module.

**Functioning modes.** If mobility is taken as an example of an attribute, one can distinguish various migration policies:
- Navigation according to a route (the site of arrival is different from the starting site)
- Circular navigation (the site of arrival is the starting site)
- Navigation directed according to a certain criterion (ex: transfer to the less loaded host or to the site having the greatest free disk space)
- Random navigation (the next site to reach is chosen at random).

*Circular navigation* and *navigation according to a route* consist of several elementary hops. Other operating modes require only one transfer: *directed navigation* and *random navigation* are mono-hop strategies.

Broadly speaking, the set of policies can be *static* or *dynamic*. It is *static* if the number or the nature of an operating mode cannot be modified after the agent creation. It is *dynamic* in the contrary case. To simplify, we considered only static sets of policies. Indeed, the implementation of a dynamic set of policies can induce a substantial modification of the choice module [11].

**Choice Module.** The decision-making process used to choose a policy does not obligatorily imply a possibility of reasoning, rational choice or cognitive structure related to a particular model of intelligence. Indeed, the module of choice splits up into two parts: *a deterministic component* and *a nondeterministic component*.

Provided with inputs and outputs, we will say that a component is deterministic if to the same inputs always correspond the same outputs. It is not deterministic if at two different moments T and T ', the same inputs can produce different outputs. The terms of inputs and outputs were selected for simplicity reasons and must be understood in our context in their most general meaning (internal or external conditions, stimuli, states, etc).

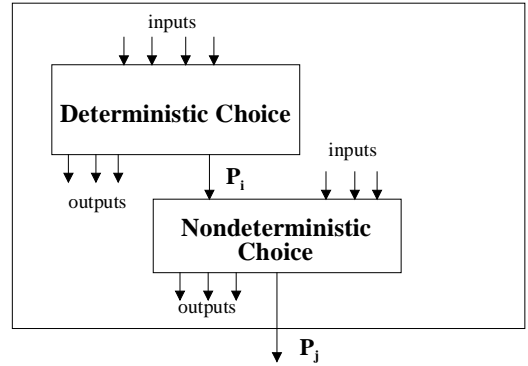

**Figure 1. Choice Module.**

The *deterministic* part of the choice module is provided with $N$ functioning policies $P_i$ ($1 \leq i \leq N$, $N > 1$). With the mobility example, migration policies could be the ones previously presented.

The *nondeterministic* part includes an additional policy $P_0$ corresponding to the *empty policy*. It means that no policy is triggered by the agent. It models the situation where, although the choice module of an attribute A was activated, no policy of A is started: there is an inhibition of the policy resulting from the deterministic choice.

That means that after the activation of the nondeterministic part of the choice module, the policy $P_j$ ($0 \leq j \leq N$) finally elected can be different from the policy $P_i$ previously selected by the deterministic component.

Each part of the choice module has its own utility. The presence of the deterministic module of choice provides the agent with a coherent behavior (rational choice) directed by the conditions of inputs. The nondeterministic module of choice ensures the agent the decoupling between its autonomy and its monitoring. It protects the agent from all recurring forms of explicit or implicit command.

If the agent included only the deterministic choice, it would be a simple program; if the nondeterministic choice were only present, the agent would not be autonomous because lacking in rationality and condemned to a chaotic behavior.

It is to be noticed that within an agent, it is neither obligatory nor necessary that the two components of the choice module of an attribute must be spatially adjacent: according to the general architecture of the agent and the properties it implements, the deterministic choice can be a part of another component of the agent (for example, when several deterministic parts of choice modules of different attributes are located at the same place). Consequently, Figure 1 must be understood as an abstract representation of a choice module.

## 4.2. An implementation example

We have implemented a software agent immersed into a real environment (network of Linux systems), an agent autonomous with regard to the mobility and having to carry out the same task on each visited site. The purpose of this application was multiple:
- To characterize in a fine way the various properties (qualities or attributes) present in the agent
- To define and implement several migration policies
- To study the realization of a complete choice module (i.e. containing deterministic and nondeterministic components) as well as its relations with the defined policies
- To verify that the absence of the agent migration control by a user did not prevent the application from functioning correctly
- To evaluate the potential of extension of this autonomy model.

**Agent task.** The task carried out by the agent on each visited site consists in collecting the users' names locally logged. Before finishing its execution, the agent transmits to the user by electronic mail the result of its various displacements, this result including the temporarily inaccessible sites (stopped machines), the sites prohibited to the agent (it does not have the necessary authorizations or the site presents an inappropriate execution environment).

**Navigation policies.** The policies at the disposal of the agent are the *random transfer* ($P_1$) and the *circular navigation* ($P_2$). These moving policies were selected for the following reasons:
- The *random transfer* offers the agent to randomly draw the name of the next site to be visited. This policy was introduced in order to differentiate it from the nondeterministic choice: in the first case, the random draw relates to the name of the next site to reach (i.e. a parameter of the policy), in the second case, the drawing of a policy among N (with N=2).
- The *circular navigation* requires the implementation of a multi-hop mechanism which implies the construction of a route and to keep up to date the list of the sites to be visited. In this configuration of navigation, the follow-up of the agent is more complex because the "umbilical link" between the agent and its launching site is broken. Lastly, it makes it possible to implement the stop of a multi-hop policy during its execution after the agent choice.

**Choice module**. The choice module breaks up into two strictly ordered contiguous levels and is materialized by the *autonomous_choice* function (written in a script language). For simplicity reasons in the design and in the functioning, the set of the navigation policies and the choice module are static: no policy is removed or added during the execution of the agent and the choice module (deterministic component and nondeterministic component) remains fixed during the lifespan of the agent.

The function *autonomous_choice* sequentially calls two functions, *deterministic_choice* and *nondeterministic_choice*:

```
function autonomous_choice
{
    deterministic_choice
    nondeterministic_choice
}
```

The *deterministic_choice* function selects or maintains the current navigation policy. There is a selection of a new policy when the preceding policy has finished, otherwise the current policy is maintained. After execution of this function, the policy $P_i$ ($1 \leq i \leq 2$) is positioned (*random* or *circular*). It will be maintained or modified by the nondeterministic part.

The nondeterministic part is implemented by the *nondeterministic_choice* function which provides a random choice among the three following possibilities: it preserves the policy fixed by the deterministic component (with the probability $Pr_1$) or it forces the execution of the *random* policy ($Pr_2$) or starts the *empty* policy $P_0$ ($Pr_3$). This one means that no navigation policy is activated. In our application, $P_0$ causes the end of the execution of the agent. Indeed, as the agent's activity mainly consists in carrying out a service (a task) on the local site then to move, it was decided that when the empty policy would be started it would cause the stopping of the agent on its local site. In an application where the agent would be richer in properties, the empty navigation policy would allow the agent to continue its execution without moving, the future of the agent depending on the properties it integrates.

**Qualities and attributes of the agent.** As well as autonomy (quality) and mobility (attribute), the agent integrates two other attributes, these two being relative to *perception*. The first is the perception by the agent of a clone of itself, i.e. a residual "incarnation" of the agent. The presence of this type of entity indicates a dysfunction of the application (and not the presence of a *replication* attribute into the agent). The second attribute is the site perception.

The agent is autonomous with regard to the mobility and non autonomous with regard to the other two attributes relative to perception because for each one of them, the agent integrates only one use policy and no choice module.

### 4.3. Possible extensions

The work undertaken on autonomy with regard to an attribute offers several tracks to be explored or deepened. We will mention four of them:

- To apply the model of autonomy to an attribute other than mobility. A candidate attribute already mentioned is the *replication* of the agent. It is also possible to build an agent autonomous with regard to the task it has to carry out

- For a given attribute, to make the set of policies eligible by the choice module dynamic. That induces a more or less major modification of the two components of this module. If there is integration within the agent of one or more policies which are by design external to it, it can be interesting to provide the agent with a software architecture which makes it *physically open* [11]

- To design an agent autonomous with regard to two different attributes. There are then two deterministic components and two nondeterministic components. A meticulous study of the interactions between the four components will have then to be undertaken. The problems to be resolved seem similar to those posed by the fitting or the organization into a hierarchy of autonomous subsystems within a broader entity, problems already identified in the biological and social fields [8]

- To integrate in the agent various models of autonomy: autonomy with regard to an attribute and, for example, a model of social autonomy.

### 5. Conclusion

As a quality, autonomy is a complex property, difficult to encircle. Nevertheless, it is possible to distinguish if an agent is or is not autonomous with regard to a particular attribute. By being nonsocial, the suggested autonomy model spares the introduction of an additional source of complexity in the guise of a second agent and with it, the complexity carried by the interactions which it can produce with the first agent. By being partial, it avoids contradictory interpretations: an agent is autonomous or non autonomous with regard to an attribute. Lastly, the model being built on the concepts of choice and uses policies, entities such as a thermostat or the *xbiff* daemon cannot be considered within this model as autonomous agents. Although provided with many assets, this model (like the others) does not exhaust all the richness of autonomy when it is considered as a *quality*.

However, combining at the same time rational choice and non determinism, this model offers a conceptual framework making it possible to think how to implement a sort of freedom of an artificial agent.